\documentclass[a4paper,fleqn]{cas-sc}

\usepackage[authoryear]{natbib}
\usepackage{textcomp}
\usepackage{amsmath}
\def\tsc#1{\csdef{#1}{\textsc{\lowercase{#1}}\xspace}}
\tsc{WGM}
\tsc{QE}

\begin{document}
\let\WriteBookmarks\relax
\def\floatpagepagefraction{1}
\def\textpagefraction{.001}

\shorttitle{Magnetoacoustic waves and high-frequency acoustic halos} 

\shortauthors{Hirdesh Kumar et al}
\title [mode = title]{A study of the propagation of magnetoacoustic waves in small-scale magnetic fields using solar photospheric and chromospheric Dopplergrams: HMI/SDO and MAST observations}

\author[1,2]{Hirdesh Kumar}[type=editor,
                        auid=000,bioid=1,
                        orcid=0000-0001-9631-1230]
\cormark[1]

\ead{hirdesh@prl.res.in}

\affiliation[1]{organization={Udaipur
Solar Observatory, Physical Research Laboratory},
            addressline={Dewali, Badi Road}, 
            city={Udaipur},
            postcode={313001}, 
            state={Rajasthan},
            country={India}}

\affiliation[2]{organization={Department of physics, Indian Institute of Technology Gandhinagar}, 
            city={Gandhinagar},
            postcode={382355}, 
            state={Gujarat},
            country={India}}

\author[1]{Brajesh Kumar}
\author[3,4]{S. P. Rajaguru}
\author[1]{Shibu K. Mathew}
\author[1]{Ankala Raja Bayanna}

\affiliation[3]{organization={Indian Institute of Astrophysics}, 
            city={Bangalore},
            postcode={34}, 
            country={India}}

\affiliation[4]{organization={Solar Observatories Group, Department of Physics and W.W. Hansen Experimental Physics Lab, Stanford University}, 
            city={Stanford CA},
            postcode={94305-4085}, 
            country={USA}}
\cortext[1]{Corresponding author}

\begin{abstract}
In this work, we present a study of the propagation of low-frequency
 magneto-acoustic waves into the solar chromosphere within small-scale
inclined magnetic fields over a quiet-magnetic
network region utilizing near-simultaneous photospheric and
chromospheric Dopplergrams obtained from the HMI instrument onboard
SDO spacecraft and the Multi-Application Solar Telescope (MAST)
operational at the Udaipur Solar Observatory, respectively.
 Acoustic waves are stochastically excited inside the convection zone of
the Sun
and intermittently interact with the background magnetic fields
resulting into episodic signals. In order to detect these episodic signals,
we apply the wavelet transform technique to the
photospheric and chromospheric velocity oscillations in magnetic
network regions. The wavelet power spectrum over photospheric and
chromospheric velocity signals show a one-to-one correspondence
between the presence of power in the 2.5-4 mHz band. Further, we
notice that power in the 2.5-4 mHz band is not consistently present
in the chromospheric wavelet power spectrum despite its presence in
the photospheric wavelet power spectrum. This indicates that leakage of
photospheric oscillations (2.5-4 mHz band) into the higher atmosphere is
not a continuous process. The average phase and coherence spectra estimated
from these photospheric and chromospheric velocity oscillations
illustrate the propagation of photospheric oscillations (2.5-4 mHz)
into the solar chromosphere along the inclined magnetic fields. Additionally,
chromospheric power maps estimated from the MAST Dopplergrams also
show the presence of high-frequency
acoustic halos around relatively high magnetic concentrations,
depicting the refraction of high-frequency
fast mode waves around $v_{A} \approx v_{s}$  layer in the solar atmosphere.
\end{abstract}


\maketitle

\section{Introduction}

Acoustic waves are well recognised as an agent of non-thermal energy transfer that couples the lower solar 
atmospheric layers, primarily through their interactions with and transformations by the highly structured 
magnetic fields that thread these layers. These waves are generated by turbulent convection within and near 
the top boundary layers of the convection zone \citep{1952RSPSA.211..564L, 1967SoPh....2..385S, 
1990ApJ...363..694G, 1998A&A...335..673B} and they resonate to form $p$-modes in the interior of the Sun. Propagation of these waves into the solar atmosphere is determined by the height dependent characteristic 
cut-off frequency $\nu_{ac}$, which takes the value of $\approx$ 5.2 mHz for the quiet-Sun photosphere 
\citep{1977A&A....55..239B, 2006ApJ...648L.151J} as estimated from $\omega_{ac} = 2\pi\nu_{ac} = 
{c_{s}}/{2H_{\rho}}$, where c$_{s}$ and H$_{\rho}$ are the photospheric sound speed and density scale height, 
respectively. Height evolution of wave phase in the chromospheric layers at frequencies larger than this 
cut-off of 5.2 mHz, while evanescent for smaller frequencies, is a well-observed feature in the quiet Sun. 
However, this condition is altered by the magnetic fields, which affect the propagation of these waves; when 
plasma $\beta$ is low ($\beta<$1), acoustic cut-off frequency $(\nu_{ac})$ is changed by a factor of 
$\cos\theta$ \citep{1977A&A....55..239B,2006ApJ...647L..77M,2016GMS...216..489C}, where $\theta$ is the angle 
between the magnetic field and the direction of gravity (which is normal to the solar surface). This fundamental 
effect of the inclined magnetic field has been identified as a key mechanism to tap the energetic low-frequency 
($p$-mode) acoustic waves to energise the solar atmospheric layers and was termed as ``magnetoacoustic 
portals" \citep{2006ApJ...648L.151J}. \cite{2013A&A...560A..84S, 2016ApJ...826...49S, 2019ApJ...871..155R, 2020ApJ...890...22A}, and \cite{2020A&A...642A..52A} have also investigated the propagation of acoustic and magnetoacoustic waves in different magnetic configurations in the lower solar atmosphere. There have also been studies of $p$-mode waves as drivers of impulsive dynamical phenomena that supply mass and energy to the solar atmosphere 
\citep{2004Natur.430..536D, 2005ApJ...624L..61D}.\\

In addition to that, we also observe high-frequency power enhancement (above the acoustic cut-off frequency) 
surrounding the strong magnetic structures such as sunspots, pores, and plages. These excess power 
enhancements are known as ``acoustic halos". This was first observed at the photosphere 
\citep{1992ApJ...394L..65B} as well as chromosphere \citep{1992ApJ...392..739B, 1993ApJ...415..847T} in the 
frequency range $\nu$ = 5.5--7 mHz, in typically weak to intermediate (B$_{LOS}$ $\approx$ 50--300 G) 
photospheric magnetic field regions. The observational studies of halos \citep{1998ApJ...504.1029H, 
2000ApJ...537.1086T, 2002A&A...387.1092J, 2004ApJ...613L.185F, 2007A&A...471..961M, 2007PASJ...59S.631N} 
reveal several features, a summary of which has been presented in \cite{2009A&A...506L...5K}. Moreover, there 
is no single theoretical model which can completely explain all the observed features despite having focused 
efforts \citep{2008AnGeo..26.2983K, 2008ApJ...680.1457H, 2009A&A...503..595H, 2009A&A...506L...5K}.  The majority of the work emphasizes the interaction of acoustic waves with the background 
magnetic fields from the photospheric to the chromospheric heights, with regards to power halos observed around 
the sunspots (\citet{2009A&A...506L...5K} and references therein). However, \cite{2001A&A...379.1052K}, \cite{2001ApJ...561..420M}, and \cite{2003A&A...405..769M} have investigated the oscillatons in the network and internetwork regions utililizing observations of lower solar atmosphere from Transition Region and
Coronal Explorer (TRACE; \cite{1999SoPh..187..229H}). The  physical characteristics of upward 
propagating acoustic waves are shaped by the topology of nearby magnetic fields. Numerical modelling done by 
\cite{2002ApJ...564..508R} and \cite{2003ApJ...599..626B} show that the propagation of acoustic waves in the 
solar atmosphere is affected by the magnetic canopy and plasma $\beta \approx$1 layer where mode conversion, 
transmission and reflection take place. \cite{2009A&A...506L...5K} utilizing the numerical simulations of 
magnetoacoustic wave propagation in a magneto-static sunspot model suggest that halos can be caused by the 
additional energy injected by the high-frequency fast mode waves, which are refracted in the vicinity of the 
transformation layer (where Alfven speed is equal to the sound speed) in the higher atmosphere due to rapid 
increase of the Alfven speed. \cite{2011SoPh..268..349S} identified some new
properties of acoustic halos in active regions. \cite{2013SoPh..287..107R} further explored  
different properties of high-frequency acoustic halos around active regions including possible 
signatures of wave refraction utilizing the photospheric velocities, vector magnetograms and lower 
atmospheric intensity observations. \\

In this article, we present an analysis of leakage of low-frequency (2.5--4 mHz) acoustic waves into the 
higher solar atmospheric layers along small-scale inclined magnetic fields of a quiet magnetic network region in the form of magnetoacoustic waves  
exploiting the high-resolution photospheric velocity observations in Fe I 6173 {\AA} line from Helioseismic 
and Magnetic Imager (HMI; \citet{2012SoPh..275..229S} instrument onboard the {\em Solar Dynamics Observatory} (SDO; 
\citet{2012SoPh..275....3P}) spacecraft and chromospheric velocity estimated from Ca II 8542 {\AA} line scan 
observations obtained from the Multi-Application Solar Telescope (MAST; 
\cite{2009ASPC..405..461M,2017CSci..113..686V}.  It is to be noted that $p$-mode oscillations are stochastically excited inside the convection zone beneath the solar photosphere and intermittently interact with the background magnetic fields, resulting in episodic signals. Hence, here we employ wavelet analysis to detect these episodic signals propagating from the photospheric to the chromospheric heights using velocity observations. We also investigate the interaction of acoustic waves with 
the background magnetic fields resulting in the formation of high-frequency acoustic halos in the velocity 
power maps of a magnetic network region and their relation with the photospheric vector magnetic fields. Our 
analysis aims to provide insight into the interaction of acoustic waves with small-scale magnetic fields of a 
quiet network region, depicting different physical phenomena. In the following, Section 2 provides a detailed 
account of the observational data used in this article, Section 3 discusses the analysis procedure and 
results obtained from the investigation, and Section 4 provides a further discussion and summary of our results.

 \section{The Observational data}
 
We employ two-height velocities observed over a magnetic network region in the disc centre to investigate the 
propagation of $p$-mode waves and appearance of high-frequency power halos and their relation with the 
background magnetic fields. We use photospheric observations in Fe I 6173 {\AA} spectral line obtained from 
the HMI instrument onboard the {\em SDO} spacecraft and chromospheric line-scan observations in Ca II 8542 
{\AA} spectral line from the MAST operational at the Udaipur Solar Observatory, Udaipur, India. The following 
sub-Sections provide brief descriptions of the instruments and data reduction done for the above 
observations.
\\ 
 
\begin{figure*}
\centering
\includegraphics[scale=0.30]{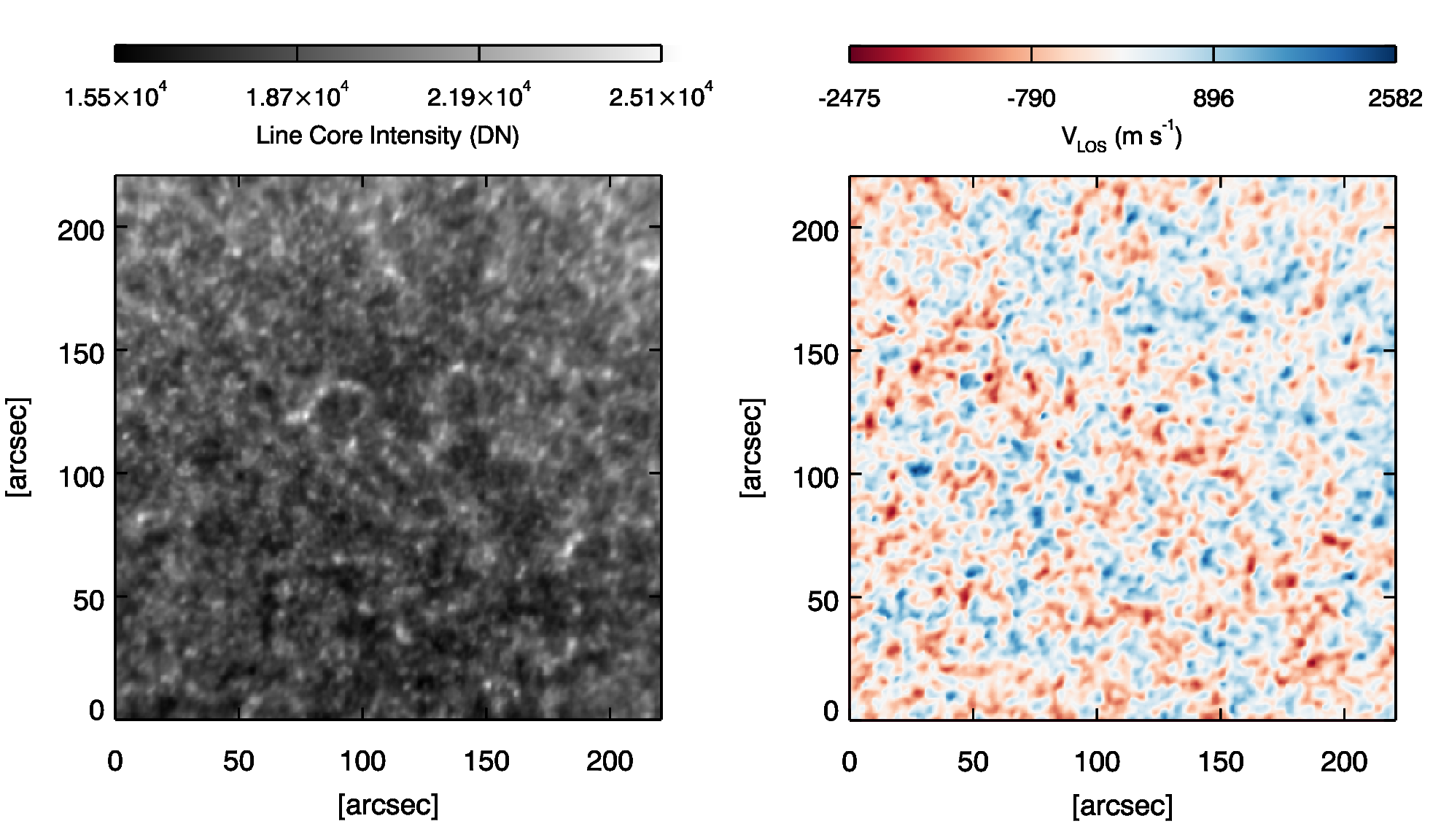}
\includegraphics[scale=0.20]{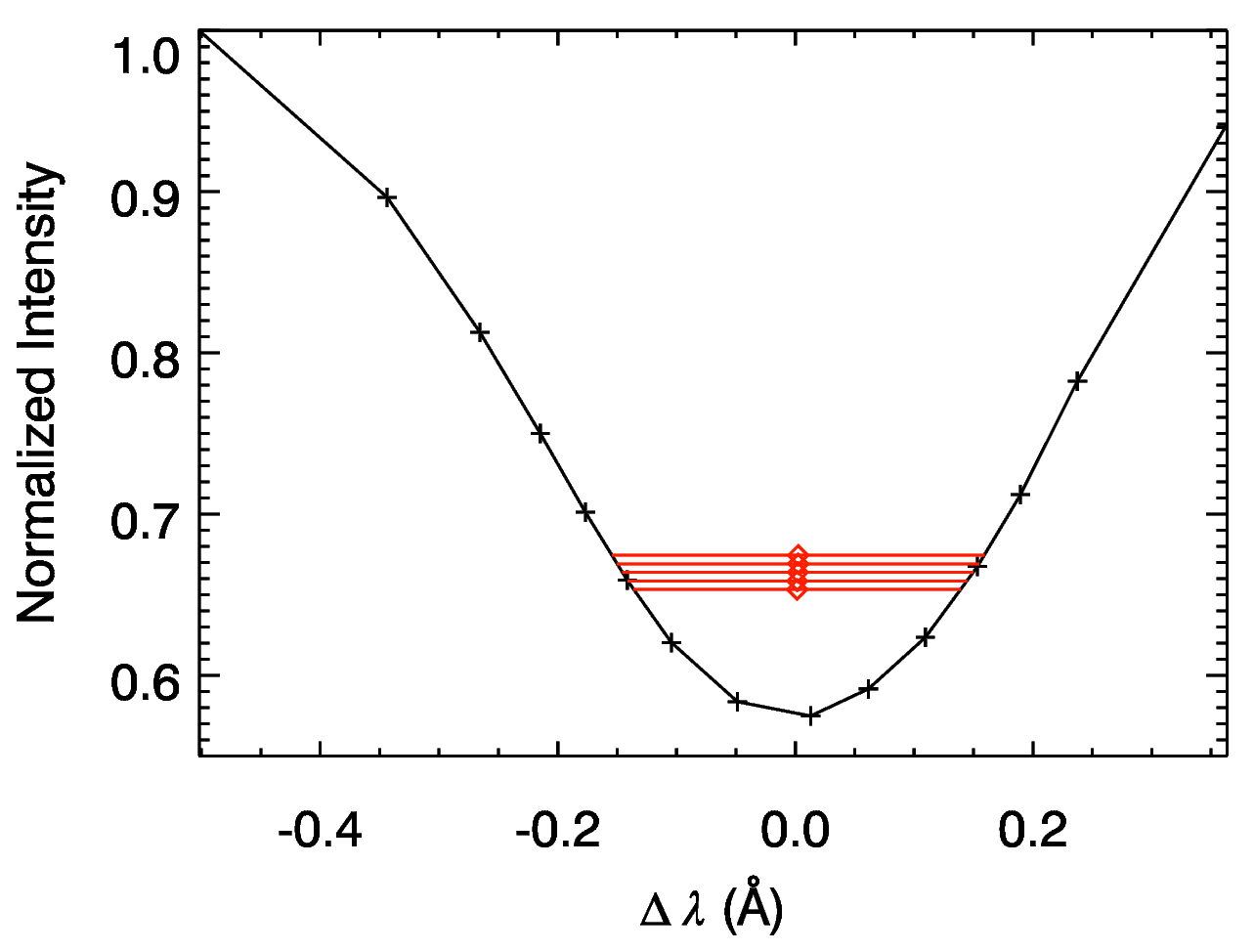}
\caption{ Sample maps of chromospheric Ca II 8542 {\AA} near line-core intensity image (\textit{top left panel}), and chromospheric 
Dopplergram (\textit{top right panel}) of a quiet magnetic network region observed on May 21, 2020 from the MAST. \textit{Bottom panel} shows a sample profile of Ca II 8542 {\AA} spectral line with five bisector points (red diamond) used to 
derive the chromospheric line-of-sight Doppler velocity. This profile is constructed from the average 
intensity over the whole field-of-view (FOV) of the MAST (c.f., \textit{top left panel}). The 15 wavelength locations were scanned in about 15 seconds.}
\label{MASTLOSDV}
\end{figure*}

\begin{figure*}
\centering
\includegraphics[scale=0.25]{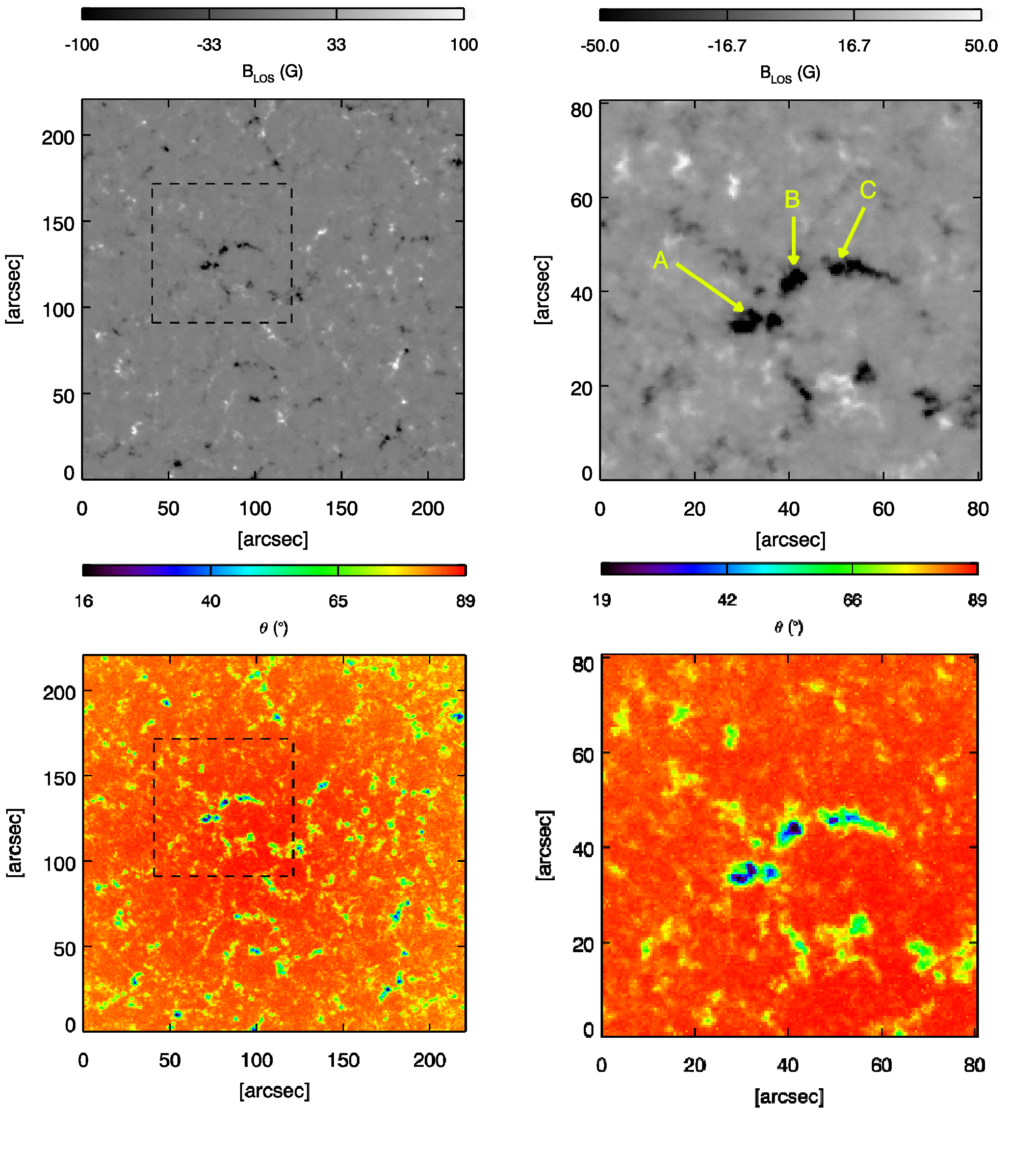}
\caption{Sample maps of average of photospheric line-of-sight magnetogram over 112 minutes duration (\textit{top left panel}), blow up region of a strong magnetic network region (\textit{top right panel}) as denoted by black dashed square box in the top left panel. Line-of-sight (LOS) magnetogram has been saturated at $\pm$100 G and $\pm$ 50 G, respectively, to bring out the small-scale magnetic features. The yellow arrows with labels $`A'$, $`B'$, and $`C'$, show the identified locations for the investigation of leakage of $p$-mode oscillations. \textit{Bottom left panel}: Sample map of $\theta$ obtained from the HMI vector magnetograms and integrated over
the duration of observations. Blow up region illustrating $\theta$ of a selected strong
magnetic network location (\textit{bottom right panel}) as denoted by black dashed square box in the left
panel.}
\label{HMI_BLOS}
\end{figure*}

\subsection{Chromospheric observations in Ca II 8542 {\AA} line from the MAST}

The MAST is a 50 cm off-axis Gregorian solar telescope situated at the Island observatory in the middle of 
Fatehsagar lake at the Udaipur Solar Observatory, Udaipur, India. It is capable of providing observations of 
the photosphere and chromosphere of the Sun. In order to obtain these observations, an imager optimized with 
two or more wavelengths is integrated with this telescope. For this purpose, two voltage-tunable lithium 
niobate Fabry-Perot etalons along with a set of interference blocking filters have been used for developing 
the Narrow Band Imager (NBI: \cite{2014SoPh..289.4007R}; \cite{2017SoPh..292..106M}). These two etalons are 
used in tandem for photospheric observation in Fe I 6173 {\AA} spectral line. However, only one of the 
etalons is used for the chromospheric observation in Ca II 8542 {\AA} (hereafter Ca II) spectral line. The 
maximum circular field-of-view (FOV) of the MAST is 6 arcmins, and diffraction-limited resolution is 0.310 
arcsec at Fe I 6173 {\AA} wavelength. Utilizing the capabilities of NBI with the MAST, we have observed a 
magnetic network region in the disk centre in the chromosphere of the Sun. The Ca II line profile has been 
scanned at 15 different wavelength positions starting from 8541.6 {\AA} to 8542.48 {\AA} within 15 s at a spatial sampling of 0.10786 arcsec per pixel on 
May 21, 2020, from 04:52:42 - 06:45:20 UT. The core of the line profile is scanned with close sampling intervals, whereas wings are sampled at higher wavelength spacing.  Moreover, there is a delay of 19 s between two consecutive line 
scans. We obtained intensity images at different wavelength positions in a FOV of 220$\times$220 arcsec$^2$. The seeing condition was stable during the observation. The top left panel of Figure 
\ref{MASTLOSDV} shows the near-core intensity image of the Ca II line during the starting time of the observations. The bottom panel of  
Figure \ref{MASTLOSDV} shows the Ca II line profile estimated from the average intensity over the 
full FOV, where plus symbols on the line profile represent the wavelength positions where the line profile 
has been scanned with the NBI. The centre points of red horizontal lines in the core of the Ca II spectral 
profile denote the bisector points, which we have used to estimate the line-of-sight velocity. We have 
explained the estimation of chromospheric velocity from these observations in Section 3 of this paper.

\subsection{Photospheric observations in Fe I 6173 {\AA} line from HMI/SDO}

The HMI instrument onboard the {\em SDO} spacecraft observes the photosphere of the Sun in Fe I 6173 {\AA} 
absorption line at six different wavelength positions within $\pm$ 172 m{\AA} window from the centre of Fe I 
6173 {\AA} line. It uses two 4096$\times$4096 pixels CCD cameras. One of the cameras, also known as a scalar 
camera, provides full disc line depth, Dopplergrams, continuum intensity, and line-of-sight magnetic fields 
at a spatial sampling of 0.504 arcsec per pixel and a temporal cadence of 45 s. The second camera, i.e. 
vector camera, is dedicated to the measurement of vector magnetic fields of the photosphere of the Sun at the 
same spatial sampling as scalar camera. It provides full-disk vector magnetograms at a lower cadence (12 
minutes) as a standard product, although it can also be obtained at a cadence of 135 s 
\citep{2014SoPh..289.3483H}. To obtain the vector magnetic field observations, Stokes parameters (I, Q, U, V) 
are observed at six different wavelength positions from the centre of Fe I 6173 {\AA} absorption spectral 
line profile. Further, to get the information of vector magnetograms of the photosphere from these Stokes 
parameters, the HMI team utilize Very Fast Inversion of the Stokes algorithm code (VFSIV;  
\cite{2011SoPh..273..267B}) to invert the Stokes profiles. The remaining 180 deg. ambiguity in the azimuthal 
field component is resolved using the minimum-energy method as described by \cite{1994SoPh..155..235M}, and 
\cite{2009SoPh..260...83L}. Thereafter, these heliocentric spherical coordinates (B$_{r}$, B$_{\theta}$, 
B$_{\phi}$) are approximated to (B$_{z}$ , -B$_{y}$ , B$_{x}$ ) in heliographic cartesian coordinates for 
ready use in various parameter studies \citep{1990SoPh..126...21G}. We have photospheric Dopplergrams, 
line-of-sight magnetograms at 45 s and vector magnetic fields at 12 min cadence. The top left panel of Figure 
\ref{HMI_BLOS} shows the average line-of-sight magnetic field image of the photosphere over the whole 
observation period as mentioned in Section 2.1 of this paper. We have tracked all these photospheric 
observables and also removed the differential rotation velocity signals using the Snodgrass formula 
\citep{1984SoPh...94...13S} and the spacecraft velocity from the HMI Dopplergrams.

\section{Analysis and Results} 

In the following, we discuss the analysis procedure of velocities, and magnetic fields over a magnetic 
network region and results obtained from the investigation.
 
\subsection{Estimation of line-of-sight velocity from Ca II 8542 {\AA} line scan observations from MAST}
 
We resample the Ca II intensity images to match the coarser HMI spatial scale after dark and flat corrections 
on the line-scan observations obtained from NBI with the MAST. We have used the cross${\_}$corr.pro routine 
available in SolarSoftWare (SSW) to co-align each Ca II intensity image with the previous line scan image. 
The bisector method \citep{Gray1976} has been used to construct chromospheric Doppler velocity. In order to 
define the reference wavelength, we have used the bisector method on the line profile constructed from the 
average intensity over full FOV. The bottom panel of Figure \ref{MASTLOSDV} shows the Ca II line profile estimated by 
taking the average over full FOV, with bisector points of five red horizontal lines at different intensity 
levels in the line core shown by red diamond; the data points marked with plus represent the locations, where 
the line has been scanned using NBI with the MAST. According to \cite{Cauzzi2008}, the wings of the Ca II 
spectral line that are 0.5 -- 0.7 {\AA} away from the line centre map the middle photosphere around $h$ = 200 
-- 300 km above the $\tau_{500} = 1$ (unity optical depth in the 500 nm continuum) whereas the line core is 
formed in the height range $h$ = 1300 -- 1500 km. Here, we have used bisector in between $\pm$ 0.14 -- 0.18 
{\AA} from the line centre. The contribution in this particular wavelength range comes from a height around 
$h$ $\approx$ 900 -- 1100 km (mean height $\approx$ 1000 km). By taking the average of five bisector points, 
we obtained a map of $\delta \lambda$, which, after a wavelength shift correction, is then used to estimate 
the line-of-sight velocity ($V_{los}$) given by the Doppler shift formula, $V_{los} = (\delta 
\lambda/\lambda_{0}) c $, where $\lambda_{0}$ is the rest wavelength, and $c$ is the speed of light. For the wavelength shift corrections, we use a diffuser to take the line-scan observations, which contains only the shift originating from the incident angle ($\theta$) of light. Further, we utilize bisector method to construct map of $\delta \lambda_{shift}$, which is subtracted from $\delta \lambda$ maps. 
Following this, by combining the time taken for line-scan and delay time, we obtain chromospheric Doppler 
velocity at a cadence of 34 s. Further, we linearly interpolate over time to match the lower cadence (45 s) 
photospheric Dopplergrams from the HMI instrument. The interpolated time series (chromospheric Dopplergrams) are now simultaneous to the photospheric Dopplergrams. The top right panel of Figure \ref{MASTLOSDV} shows the 
chromospheric Doppler velocity constructed from MAST observations. For aligning the photospheric and 
chromospheric observations, we use 20 min averages of line core intensity of Ca II spectral profile and 
line-of-sight magnetograms and identify similar features for correlation tracking to get sub-pixel 
accuracy.\\

\subsection{Photospheric Observables from HMI instrument}

We use photospheric Dopplergrams obtained from the HMI instrument, which sample Fe I 6173 {\AA} spectral 
line. \cite{2006SoPh..239...69N} investigated formation process of Fe I 6173 {\AA} line profile and derive a 
height range of $h$ = 20 -- 300 km above the continuum optical depth ($\tau_{c}$ = 1, which corresponds to z 
= 0) whereas line continuum is formed around $h$ = 20 km above z = 0. \cite{2011SoPh..271...27F} used 3d 
time-dependent radiation hydrodynamic simulation to produce Fe I 6173 {\AA} line profile and calculated 
Doppler velocity to compare with the observations derived with the HMI instrument. They concluded that 
$V_{HMI}$ forms at a mean height of 150 km. Other photospheric observables used in this study are B$_{LOS}$ 
and B = [B$_{x}$, B$_{y}$, B$_{z}$] at a cadence of 45 s and 12 min, respectively. The magnetic fields of the 
quiet Sun rapidly evolve over a short time duration and decay \citep{2019LRSP...16....1B}. To examine 
the role of small scale magnetic fields on the leakage of $p$-mode waves and formation of high-frequency 
halos, we have taken the average of B$_{LOS}$ over the whole observation duration as shown in the top left 
panel of Figure \ref{HMI_BLOS}. The average B$_{LOS}$ map shows only the permanent magnetic features. It has been saturated between $\pm$ 100 G to bring out small scale magnetic 
patches, and a strong network region is selected for final analysis, which is demarcated with a black dashed 
square box. The top right panel of Figure \ref{HMI_BLOS} shows the blow-up region of B$_{LOS}$. From the 
vector magnetograms, we have B$_{x}$, B$_{y}$, and B$_{z}$ components of the magnetic fields. Similar to 
\cite{2016ApJ...817...45R} and \cite{2019ApJ...871..155R}, we estimate field inclination $\theta$ = $90. - (180./\pi)|$tan$^{-1}$(B$_{z}$/B$_{h}$)$|$, where B$_{z}$ is the radial component of the magnetic fields, and B$_{h}$ = 
$\sqrt{B_{x}^2+B_{y}^2}$ is horizontal magnetic fields. Map of $\theta$ averged over the whole observation period 
is shown in the bottom left panel of Figure \ref{HMI_BLOS}, where $\theta$ = 0 degrees corresponds to vertical and $\theta$ = 
90 degrees represents horizontal magnetic fields. It is to be noted that we have averaged B$_{LOS}$ and 
$\theta$ maps over the whole observation period; this may lead to an averaging out of small scale emerging 
magnetic fields in between the observation period.  Finally, we have selected three different locations ($`A$', $`B$', and $`C$') with magnetic field strength ($|B_{LOS}|$) greater than 30 G to study the propagation of $p$-mode waves from the photosphere into the 
chromosphere as shown with the yellow arrowheads in the blow-up region of B$_{LOS}$ map (c.f. top right panel of 
Figure \ref{HMI_BLOS}), which is our region of interest (ROI). In our analysis, we have considered rasters of 10$\times$10 pixels in the aforementioned identified locations on which the velocity signals are averaged in order to take into account seeing related fluctuations on spatial scales. The values of 
$<|B_{LOS}|>$, $<\theta>$ and magnetically modified cut-off frequency $\nu^{B}_{ac} = \nu_{ac} cos\theta$
over locations $`A$', $`B$', and $`C$' are listed in the Table \ref{Table_Location}.

\begin{table}
\caption{ Table represents the values of average line-of-sight magnetic fields ($<|B_{LOS}|>$), inclination ($<\theta>$) and cut-off 
	frequency ($\nu^{B}_{ac}$) over selected locations.}
\begin{tabular}{|p{1.5cm}|p{1.5cm}|p{1.5cm}|p{1.5cm}|}
 \hline
	Location& $<|B_{LOS}|>$ (G) & $<\theta>$ (degree) & $\nu^{B}_{ac} = \nu_{ac}cos\theta$ (mHz)\\
 \hline
 $A$ & 42.9 &  64.2 & 2.3 \\
 $B$    &68.7 & 58.1 & 2.7 \\
 $C$ &   38.5  & 62.7 & 2.4 \\
  \hline
\end{tabular}
\label{Table_Location}
\end{table}

\subsection{Wavelet Analysis of HMI and MAST Velocity Data}

 In order to detect the presence of episodic wave propagation signals, from the photosphere to the chromosphere through the intermittent interactions between $p$-modes and the magnetic fields, we have constructed the wavelet power spectrum of the velocity oscillations at the identified locations $`A$', $`B$', and $`C$'.

\begin{figure}[h!]
\centering
\includegraphics[scale=0.26]{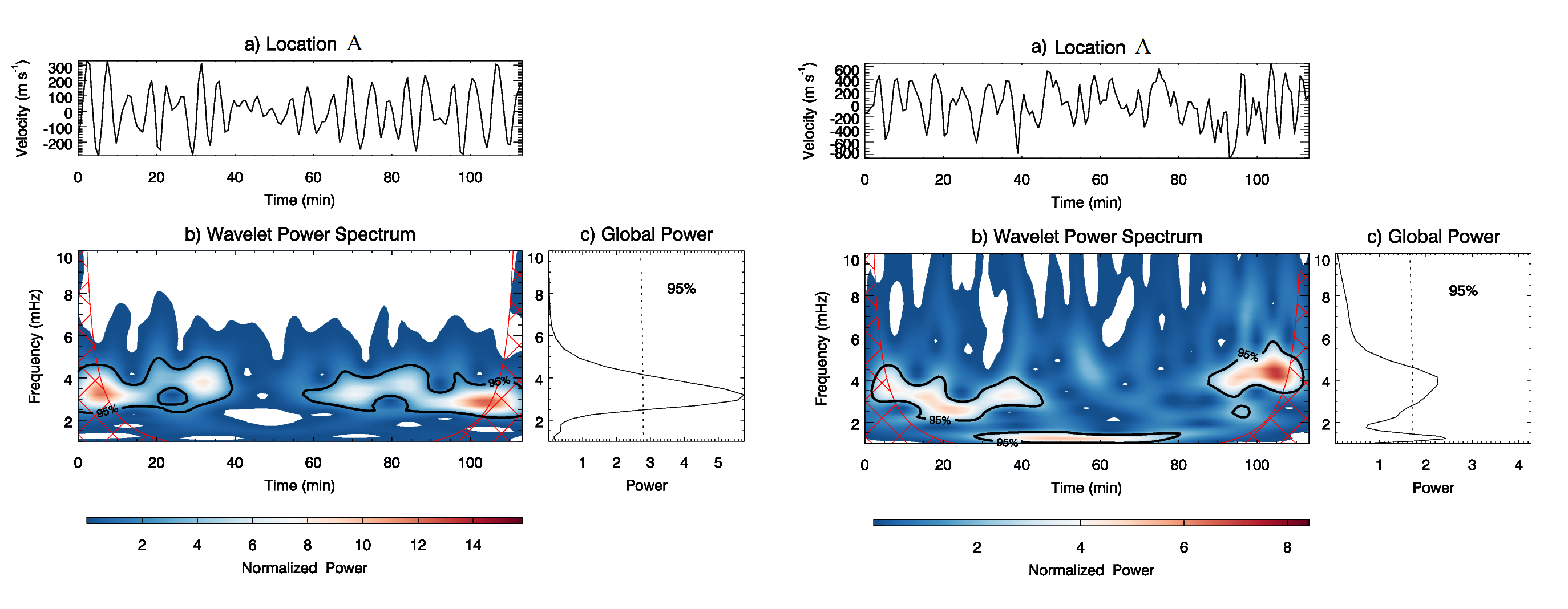}
\includegraphics[scale=0.26]{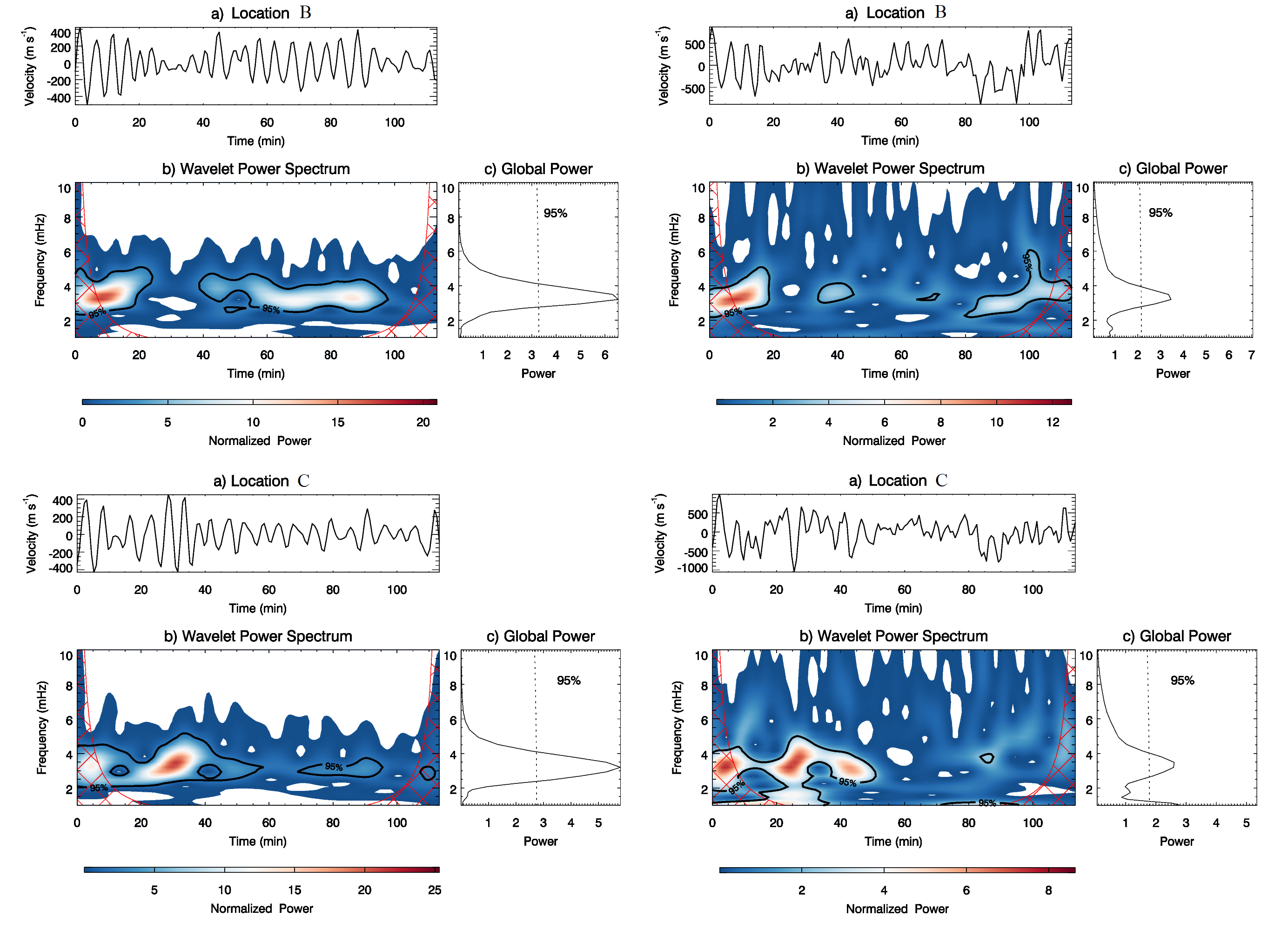}
\caption{\textit{Top left panel}: Upper panel (a) shows the temporal evolution of average velocity signals 
obtained from the photospheric Dopplergrams for the location $`A$'. Bottom panel shows the WPS and the GWPS in 
(b) and (c), respectively, computed from velocity time series. In the WPS, the solid lines demonstrate 
regions with the 95$\%$ confidence level, and the hatched region indicates the cone of influence. The colour 
scale represents the wavelet power. The dotted line in GWPS shows a confidence level at 95$\%$. \textit{Top right 
panel}: Same as \textit{top left panel}, but from chromospheric velocity for location $`A$' as estimated from MAST 
observations. {\textit{Middle and bottom panels} are for the locations $`B$' and $`C$', respectively.}}
\label{HMI_MAST_Wavelet_Power_ABC}
\end{figure}

\begin{figure*}[h!]
\centering
\includegraphics[scale=0.22]{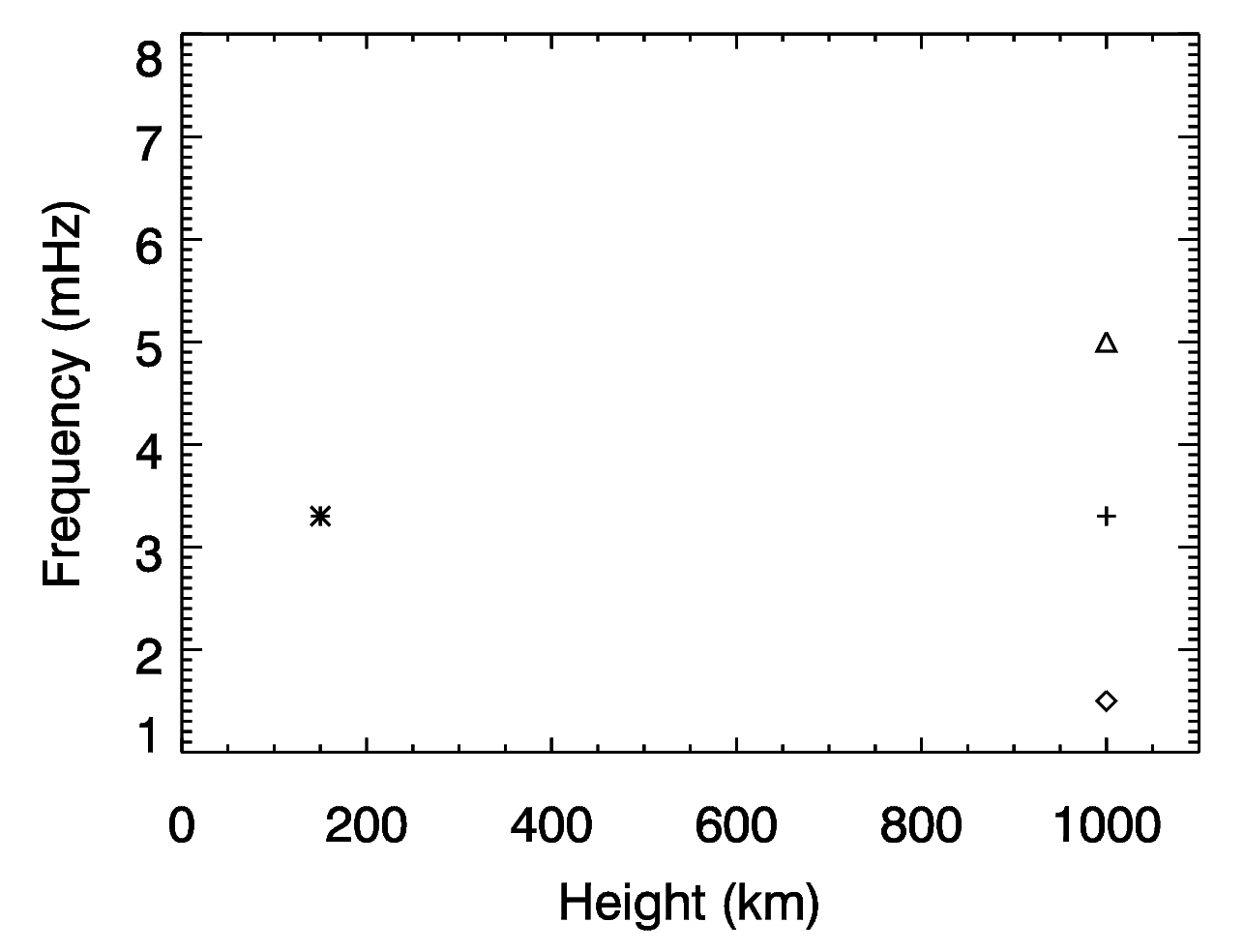}
\caption{Plot shows the frequency versus height of the oscillations
observed in the photospheric and chromospheric wavelet power spectrum. In this plot, the asterisk shows the dominant photospheric frequency observed at 3.3 mHz,
whereas the diamond, plus, and triangle represent the chromospheric
frequencies observed at 1.5, 3.3, and 5.0 mHz, respectively. The presence
of 3.3 mHz oscillations in the chromosphere is associated with the
leakage of underlying photospheric oscillations along the inclined
background magnetic fields.}
\label{Freq Height}
\end{figure*}

Wavelet analysis is a powerful technique to investigate non-stationary time series or where we expect localized power variations. Thus, to 
determine the period of episodic signals present in the identified regions, we  apply the wavelet 
technique \citep{1998BAMS...79...61T} on the photospheric and chromospheric velocity time series of locations $`A$', $`B$', and $`C$' obtained from HMI and MAST observations. \cite{1998BAMS...79...61T} 
presented a detailed description of the methodology used as the basis for this study. Here, we use the Morlet wavelet, a product of a Gaussian function and a sine wave. In the wavelet power 
spectrum (WPS), we limit our investigation to a region inside a ``cone of influence" corresponding to the 
periods of less than 25 $\%$ of time series length. We have also overplotted the confidence level at 95 $\%$ as 
shown by a solid black line in WPS. For the different locations on the photospheric velocities, we note the 
presence of significant power around the 2.5--4 mHz band in WPS, which is associated with the $p$-mode 
oscillations (c.f. left panels of Figures \ref{HMI_MAST_Wavelet_Power_ABC}). The WPS obtained from chromospheric velocities at identified locations 
also shows the power around 2.5--4 mHz band and above 5.5 mHz (c.f. right panel of Figures 
\ref{HMI_MAST_Wavelet_Power_ABC}). 
Further, the WPS is collapsed over the whole observation time to get the Global Wavelet power spectrum 
(GWPS). If power is present during the whole length of observation time in WPS, it would also be reflected in 
the GWPS. The GWPS is similar to the commonly used Fourier power spectrum. In Figures 
\ref{HMI_MAST_Wavelet_Power_ABC} GWPS 
show the peak around 3 mHz above the confidence level in photospheric and chromospheric locations. 
 Importantly, we note that WPS constructed from the chromospheric velocity  time series reveals the presence of 
significant power around the 3 mHz band at the same time as in the WPS obtained from the photospheric velocity 
time series (c.f. Figures \ref{HMI_MAST_Wavelet_Power_ABC}). For instance, significant power is present in the WPS constructed from HMI 
velocity of location $`A$' in 0--40 minutes duration in 2.5--4 mHz band (c.f. top left panel of Figure 
\ref{HMI_MAST_Wavelet_Power_ABC}), and in the same time interval, we notice significant power in WPS obtained 
from the chromospheric velocity time series (c.f. top right panel of Figure \ref{HMI_MAST_Wavelet_Power_ABC}). We have 
found similar results in other selected locations (c.f. Figure \ref{HMI_MAST_Wavelet_Power_ABC}).  We also notice that despite the presence of significant power in photospheric WPS, we do not see the power in chromospheric WPS. For example prominent power is present in the location $`B'$ in 40--100 minute time duration in photospheric WPS (c.f. middle left panel of Figure \ref{HMI_MAST_Wavelet_Power_ABC}), whereas feeble power is observed in the chromospheric WPS (c.f. middle right panel of Figure \ref{HMI_MAST_Wavelet_Power_ABC}), pointing that low-frequency acoustic waves intermittently interect with background magnetic fields. We also plot frequency versus height (c.f. Figure \ref{Freq Height}) of the oscillations observed in the photospheric and chromospheric wavelet power spectrum obtained at two different heights in the solar atmosphere. In the photospheric WPS, we see the presence of dominant 5-min (3.3 mHz) oscillations, however chromospheric WPS indicate the presence of 1.5, 3.3, and 5.0 mHz oscillations. The presence of 3.3 mHz oscillations in the chromosphere is assocaited with underlying photospheric oscillations, whereas 1.5 and 5.0 mHz oscillations are chromospheric in nature. These findings are consistent with earlier numerical simulations done by \cite{2019A&A...623A..62K}, and \cite{2023MNRAS.518.4991K}, where they have derived the cutoff frequency of upward propagating magnetoacoustic waves. Moreover, the estimated acoustic cut-off frequency 
($\nu^{B}_{ac}$) (c.f. Table \ref{Table_Location}) in the locations mentioned above shows that the quiet-Sun 
photospheric acoustic cut-off frequency ($\nu_{ac}$) is adequately reduced in the presence of inclined magnetic 
fields. Thus, it indicates that the power present around 2.5--4 mHz band in the  chromospheric WPS is possibly associated with the leakage of the photospheric $p$-mode oscillations into the 
chromosphere.

 \subsection{Phase and coherence spectra}

Wavelet power spectrums of locations $`A$', $`B$', and $`C$' of chromospheric Doppler velocities show one-to-one correspondence in the $\nu$ = 2.5--4 mHz band with the photospheric 2.5--4 mHz band, suggesting that chromospheric power in 2.5--4 mHz band are related to the leakage of photospheric $p$-mode oscillations. Hence, we estimate the phase and coherence spectra over the region of interest (ROI) (c.f. right panel of Figure \ref{HMI_BLOS}). We would like to mention here that, spatial maps of phase and coherence do not show clear one-to-one correlations. It might be due to the larger height difference between the photospheric and the chromospheric Dopplergrams, and also due to non-achievement of sub-pixel accuracy in the alignment of the HMI and the MAST observations due to seeing fluctuations. Hence, we have not included these maps in our results.  Here, we present the average phase and coherence over the ROI. In order to estimate the phase and coherence between two signals at each pixel, we calculate the cross-spectrum of two evenly sample time series ($I_{1}$ and $I_{2}$) in the following way:\\

\begin{equation}
\centering
X_{12}(\nu) =  I_{1}(\nu)\times I^{*}_{2}(\nu) 
\end{equation}

Where I's are the Fourier transforms, and $*$ denotes the complex conjugate. The phase difference between two-time series is estimated from the phase of complex cross product ($X_{12}(\nu)$)\\

\begin{equation}
\centering
\delta\phi(\nu) = tan^{-1}(Im(<X_{12}(\nu)>)R(<X_{12}(\nu)>))
\end{equation}  

We are adopting the convention that positive $\delta\phi(\nu)$ means that signal 1 leads 2, i.e., a wave is propagating from lower height to upper height and vice-versa. Further, the magnitude of $X_{12}(\nu)$ is used to estimate the coherence (C), which ranges between 0 to 1. 

\begin{equation}
\centering
C(\nu) = sqrt(|<X_{12}(\nu)>|^{2} / <|I_{1}|^{2}> <|I_{2}|^{2}>)
\end{equation} 

Where $<.>$ denotes the average. Coherence is a measure of the linear correlation between two signals, 0 indicates no correlation and 1 means perfect correlation. In our case, we have averaged over segments of length 76 data points, which correspond to 57 minutes at a cadence of 45 seconds. We assume that this is the better way to estimate the $\delta\phi(\nu)$ and  $C(\nu)$ due to the intermittent nature of the interaction of acoustic waves with the background magnetic fields \citep{2019ApJ...871..155R}.\\

\begin{figure*}[h!]
\centering
\includegraphics[scale=0.22]{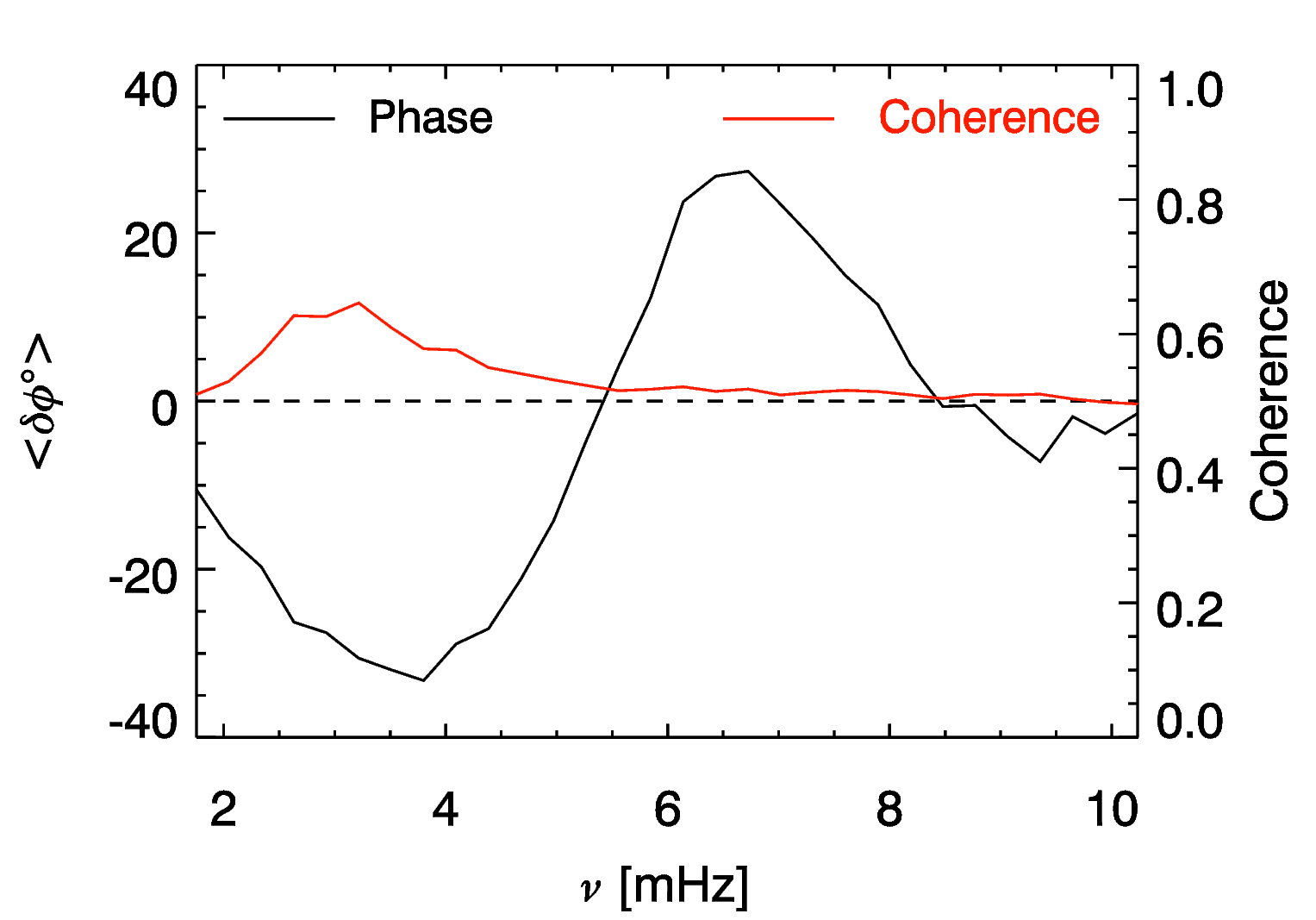}
\caption{ Spatially averaged phase and coherence spectra estimated from photospheric and chromosphere Doppler velocities over region of interest (ROI)  as shown in the right panels of Figure \ref{HMI_BLOS}.}
\label{HMI_MAST_Phase_Coherence}
\end{figure*}

Figure \ref{HMI_MAST_Phase_Coherence} shows the average phase (black) and coherence (red) spectra between photospheric and chromospheric Doppler velocity observations estimated over the region of interest (ROI) as shown by the black dashed square box on the left panels of Figure \ref{HMI_BLOS}, and blow up region of the same is depicted in the right panels of Figure \ref{HMI_BLOS}. In between the photosphere and the chromosphere, atmospheric gravity waves also contribute to the $\delta\phi$. Gravity waves show the negative phase propagation while transporting energy in the upward direction in the lower frequency band in $k-\omega$ diagram \citep{2008ApJ...681L.125S}. In our analysis, we also found the negative $\delta\phi$ in the phase estimation from the photospheric and the chromospheric velocities. This is possibly associated with the contribution of gravity waves. However, we found that $\delta\phi$ started increasing around $\nu$ = 3.5 mHz and continue up to $\nu$ = 6.5 mHz after which it decreases. The decline of $\delta\phi$ after 6 mHz is possibly associated with steepening of acoustic waves into the shocks. Moreover, the coherence is above 0.5 up to 6 mHz and then slightly decreases. Overall, the estimated average phase and coherence spectra from photospheric and chromospheric velocities indicate the propagation of $p$-mode waves and are qualitatively in agreement with phase travel time estimated by \citep{2019ApJ...871..155R} utilizing the photospheric and lower chromospheric intensity observations.

\subsection{Photospheric and chromospheric behaviour of high-frequency acoustic halos}   

\begin{figure*}[ht!]
\centering
\includegraphics[scale=0.21]{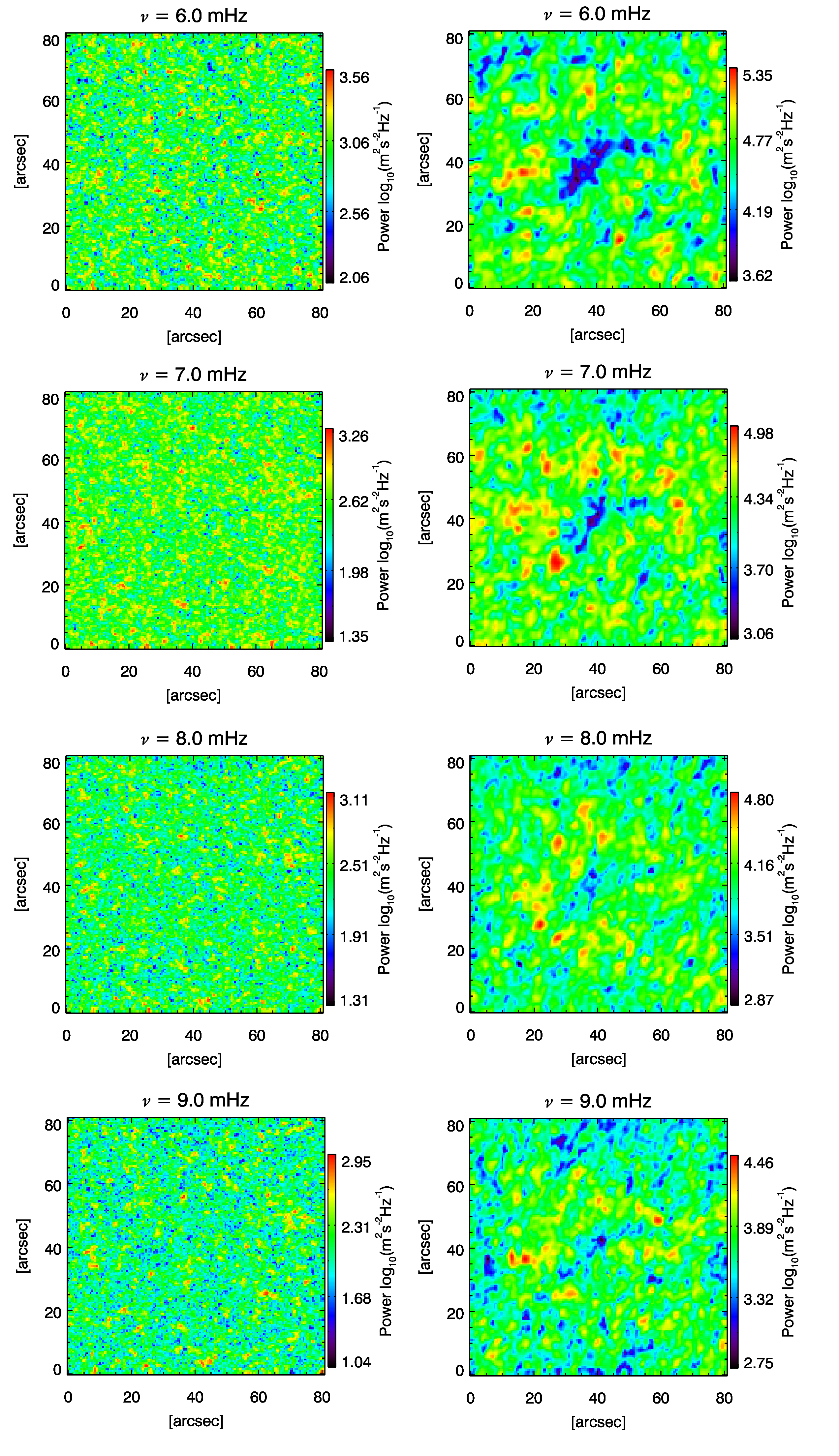}
\caption{\textit{Left panel}: Power maps at different frequencies constructed from photospheric velocity 
obtained from HMI instrument over a small strong magnetic network region as shown  in the right panels of Figure \ref{HMI_BLOS}. \textit{Right panel}: Same as \textit{left panel} 
but from chromospheric velocity estimated from MAST observations.}
\label{HMI_MAST_Power_maps}
\end{figure*}

\begin{figure*}[ht!]
\centering
\includegraphics[scale=0.24]{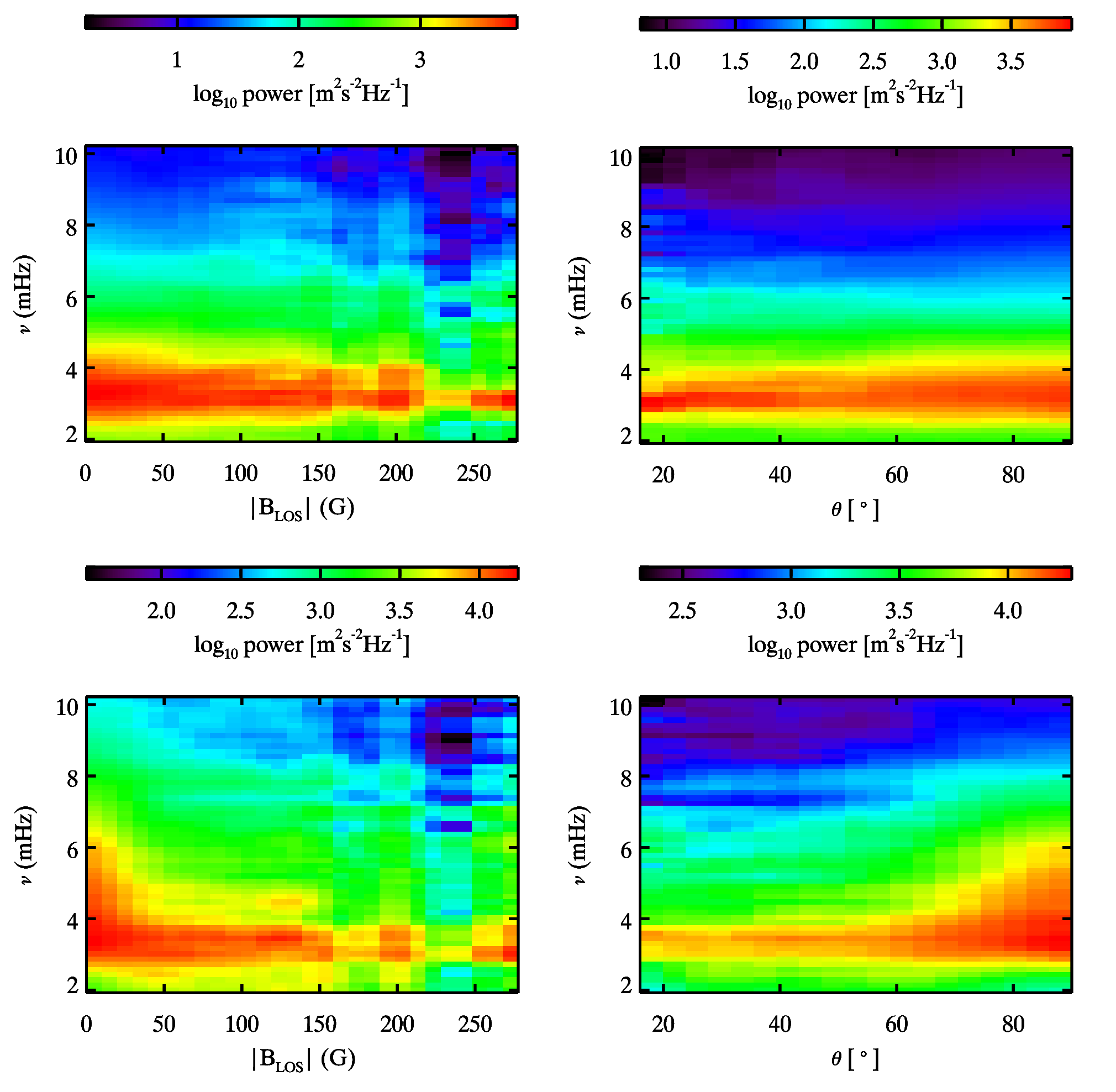}
\caption{\textit{Top panel}: Photospheric power are averaged over photospheric magnetic fields, 10 G bins in  $|B_{LOS}|$ (\textit{left hand side}) and 4 degree bins in $\theta$ (\textit{right hand side}), respectively, over a magnetic network region as shown in the right panels of Figure \ref{HMI_BLOS}. \textit{Bottom panel}: Same as \textit{top panel}, but estimated from chromospheric power over photospheric magnetic fields ($|B_{LOS}|$ and $\theta$), respectively.}
\label{HMI_MAST_PowerMaps_Blos_Theta_Small_FOV}
\end{figure*}

The propagation of acoustic waves into the higher solar atmospheric layers in magnetized regions results into 
different physical phenomena. One of them is the formation of high-frequency acoustic halos. To gain an 
insight into the interaction of acoustic waves with the background magnetic fields in the quiet magnetic 
network region (where the magnetic field strength is typically small), we have constructed power maps in 
different frequency bands, i.e. $\nu$ = 6, 7, 8, and 9 mHz from photospheric and chromospheric velocities 
over a strong magnetic network region as indicated in the top left panel of Figure  \ref{HMI_BLOS} by the 
black dashed square box and blow up region of the same is shown here in the top right panel. These maps are integrated over $\pm$ 0.5 mHz frequency band around the center as 
frequency mentioned on the top of each maps (c.f. Figure \ref{HMI_MAST_Power_maps}). Photospheric and 
chromospheric power maps are shown in the left and right panels of Figure \ref{HMI_MAST_Power_maps}, 
respectively. We do not see high frequency power halos in the photospheric power maps (c.f. left panels of 
Figure \ref{HMI_MAST_Power_maps}) surrounding strong network region. However, chromospheric power maps show 
high frequency acoustic halos around the magnetic network region at $\nu$ = 6, 7, 8, and 9 mHz maps as shown 
in the right panels of Figure \ref{HMI_MAST_Power_maps}. In the centre of strong magnetic network 
concentrations, chromospheric power is suppressed (c.f. right panels of Figure \ref{HMI_MAST_Power_maps}). It is to be noted that the magnetic fields of network region spread and diverge at 
chromospheric height and become significantly inclined in higher atmosphere and changing the height of $\beta$ = 1 layer. In our analysis, significant enhancement in chromospheric power is observed in areas surrounding the strong magnetic 
network region in the form of patches. Further, to better understand the 
behaviour of high-frequency acoustic halos with respect to photospheric magnetic fields, we have averaged power over 
a bin of 10 G in $|B_{LOS}|$ and 4-degree bin in $\theta$. Photospheric and chromospheric power as a function of 
photospheric $|B_{LOS}|$ and $\theta$ (integrated over whole observation period) are shown in the top and 
bottom panels of Figure \ref{HMI_MAST_PowerMaps_Blos_Theta_Small_FOV}, respectively. It is to be noted that 
\cite{2013SoPh..287..107R} have estimated power over different ranges of magnetic field strength and 
inclination from the analysis of different active regions. However, we investigate halos in a quiet magnetic 
network region, where the magnetic field strengths are much smaller compared to active regions. Photospheric power as a function of $|B_{LOS}|$ shows that more power is present in the 2.5--4.5 mHz band in relatively small magnetic field strengths with more inclined magnetic field regions (c.f. top panels of Figure 
\ref{HMI_MAST_PowerMaps_Blos_Theta_Small_FOV}). Interestingly, chromospheric power as a function of 
photospheric $|B_{LOS}|$ and $\theta$ shows that excess power in the high-frequency ($\nu >$ 5 mHz) band is present in the 
small magnetic field strength ($|B_{LOS}|<$ 60 G) and more inclined magnetic field regions ($\theta >$ 60 
degree) (c.f. bottom panel of Figure \ref{HMI_MAST_PowerMaps_Blos_Theta_Small_FOV}).

\section{Discussion and Conclusion}

Understanding the physics of interactions between acoustic waves and magnetic fields requires
identifying and studying the observational signatures of various associated phenomena in the solar atmosphere,
viz. mode conversion, refraction and reflection of waves, formation of high-frequency acoustic halos etc.
Most of the work in this field emphasizes the propagation of acoustic waves and their interaction within the
highly magnetized regions like sunspots and pores. However, it is not well understood as whether we also expect
similar behaviour in a quiet magnetic network region.
Multi-height observations covering photospheric and chromospheric layers, especially through intensity
imaging observations, are widely used towards the above. Simultaneous velocity observations of both photospheric
and chromospheric heights are however rare. In this work, in order to address some of the above questions,
we have observed a quiet-Sun magnetic network region
in the chromospheric Ca II 8542 {\AA} line using the NBI instrument now operational at MAST/USO. Using
spectal scans of this region, we have derived chromospheric Doppler velocities. Simultaneous
photospheric Doppler velocity and vector magnetic field of this region was also extracted from SDO/HMI 
data.\\

 We employ wavelet analysis technique to investigate the intermittent interaction of low-frequency (2.5--4.0 mHz) acoustic waves with background magnetic fields in the velocity oscillations at the photospheric and the chromospheric heights using the aforementioned observations. Therefore, we have estimated wavelet power spectra in the selected locations to investigate the propagation of low-frequency acoustic waves from the solar photosphere into the  chromosphere as magnetoacoustic waves. These waves are now recognized as a potential candidate to heat the lower solar 
atmosphere. The investigation of \cite{2006ApJ...648L.151J} utilizing the velocity observations shows that low-frequency magnetoacoustic waves can propagate into the lower solar chromosphere from the 
small-scale inclined magnetic field elements. Later on, the detailed investigation done by 
\cite{2019ApJ...871..155R} provides evidence that a copious amount of energetic low-frequency acoustic waves 
channel through the small-scale inclined magnetic fields. Usually, acoustic waves of frequency less than 
photospheric cut-off frequency ($\nu_{ac} \approx$ 5.2 mHz) are trapped below the photosphere in the quiet 
Sun. However, the presence of magnetic fields alters the cut-off frequency, and it is reduced by a factor of 
cos$\theta$: $\nu^{B}_{ac} = \nu_{ac} cos\theta$ \citep{1977A&A....55..239B}.  In our analysis, we have
found a significant reduction in the acoustic cut-off frequency, following the above formula, in the presence 
of inclined magnetic fields. Thus, the presence of low-frequency acoustic waves in the chromospheric wavelet power spectra is clearly associated with the leakage of photospheric oscillations. 
Moreover, the WPS of both heights show a clear 
association of chromospheric (5-min) oscillations with the underlying photospheric oscillations as evident from the appearance of low-frequency acoustic waves in chromospheric WPSs approximately at the same time as in 
the photospheric WPS. The investigation of the average phase over the ROI (80$\times$80 arcsec$^{2}$, c.f. right panel of Figure \ref{HMI_BLOS}) also shows the turn-on positive phase around $\nu$ = 3.5 mHz and continues upto 6.5 mHz, indicating the channelling of photospheric $p$-mode oscillations into the chromosphere. Additionally, the estimated coherence is above 0.5 up to around 8 mHz, and then it decreases. Nevertheless, the bigger height difference between the two observables also reduces the coherence because these waves travel a long distance, hence losing coherency. Our results are consistent with the previous findings related to the leakage of 
$p$-mode oscillations into higher solar atmospheric layers along inclined magnetic fields 
\citep{2004Natur.430..536D, 2005ApJ...624L..61D, 2006ApJ...648L.151J, 2007A&A...461L...1V, 
2019ApJ...871..155R}. However, our results add to the previous findings by illustration of a one-to-one correspondence between oscillatory signals present in the WPS of photospheric and the chromospheric velocity observations of a quiet-Sun magnetic network region. Nevertheless, we also note that leakage of low-frequency acoustic waves along the inclined magnetic fields is not a continuous process, indicating that energy flux estimated from short duration observations could be underestimated or overestimated.\\

As regards the formation of high frequency acoustic halos surrounding magnetic concentrations, the 
information gathered from the same is also important to map the thermal and magnetic structure of different 
heights in the solar atmosphere. The theoretical investigations of \cite{2002ApJ...564..508R}, and 
\cite{2003ApJ...599..626B} discussed the effect of magnetic canopy present in the solar atmosphere and plasma 
$\beta$ $\approx$ 1 location. The plasma $\beta \approx$ 1 separates two different regions of gas and 
magnetic pressure dominance. \cite{2001SoPh..203...71G} provided a model of plasma $\beta$ variation with 
height in the solar atmosphere over a sunspot and a plage region. Figure 3 of \cite{2001SoPh..203...71G} 
shows that $\beta$ is $\approx$ 1 for plage region at a height of around 1 Mm, while for sunspot it lies 
below. It is suggested that mode conversion, transmission, and reflection of waves occur at $\beta$ $\approx$ 
1 \citep{2002ApJ...564..508R, 2003ApJ...599..626B, 2006ApJ...653..739K, 2009A&A...506L...5K, 
2012ApJ...746...68K, 2012A&A...538A..79N}. The numerical simulation of \cite{2009A&A...506L...5K} proposes 
that high-frequency fast mode waves which refract in the higher atmosphere around $\beta$ $\approx$ 1 due to 
rapid increase of the Alfven speed can also cause high-frequency acoustic halos around sunspots. Further, 
they have added that high-frequency halos should form at a distance where the refraction of fast mode 
acoustic wave occurs above the line formation layer, i.e., where the Alfven speed is equal to sound speed 
above the photosphere. Else, it would not be possible to detect these halos in observations. We have found 
high-frequency acoustic halos around a high magnetic concentrations in quiet magnetic network region in the 
chromospheric Fourier power maps (c.f. right panels of Figure \ref{HMI_MAST_Power_maps}). However, halos are 
not observed in the photospheric power maps. The absence of the acoustic halos in photospheric power maps 
indicate that formation height of photospheric velocity is far below the mode conversion layer i.e., $\beta 
\approx$ 1, as suggested by \cite{2006ApJ...653..739K} and \cite{2013SoPh..287..107R}. Therefore, the presence of halos in the chromospheric power maps is possibly associated with the 
injection of high frequency fast mode waves, which are refracting from the magnetic canopy and $\beta 
\approx$ 1 layer as proposed by \cite{2009A&A...506L...5K} and \cite{2012A&A...538A..79N}. This fact is also supported by a model of plasma $\beta$ variation in the solar atmosphere 
\citep{2001SoPh..203...71G} suggesting that $\beta \approx$ 1 at a height of $\approx$ 1000 km over a plage 
region. For a quiet magnetic network region like ours, we do expect that $\beta \approx$ 1 layer might be 
above 1000 km. The presence of high-frequency acoustic halos in chromospheric power maps is consistent with 
the explanation provided by \cite{2006ApJ...653..739K, 2009A&A...506L...5K, 2010A&A...510A..41K, 
2012A&A...538A..79N}, and \cite{2013SoPh..287..107R}. The chromospheric power maps also show the power 
deficit in the high magnetic concentrations in the quiet magnetic network region. This could be associated 
with the energy loss due to the multiple mode transformations in a flux tube \citep{1997ApJ...486L..67C}. The 
analysis of chromospheric power dependence on photospheric $|B_{LOS}|$ and $\theta$ shows that high-frequency 
power is also significantly present in the small magnetic field strength and in nearly horizontal magnetic 
field regions. This finding is qualitatively in agreement with the earlier work \citep{2011SoPh..268..349S}, 
which suggests that the excess power in high-frequency halos is present at the more horizontal and 
intermediate magnetic field strength region. Our results thus show that small scale magnetic fields also have a significant effect on the propagation 
of acoustic waves in the solar atmosphere. It is to be noted that we have utilized high resolution 
photospheric and chromospheric Doppler velocities to investigate the interaction of acoustic waves with 
the small scale magnetic fields. We intend to follow up with similar analyses using near-simultaneous multi-height velocity and magnetic 
field observations from MAST and the newer Daniel K Inouye Solar Telescope (DKIST; \cite{2020SoPh..295..172R}) along with HMI and AIA instruments on board SDO spacecraft.\\


\section*{Acknowledgements}

We acknowledge the use of data from HMI instrument onboard the {\em Solar Dynamics Observatory} spacecraft of 
NASA. This work utilizes data obtained from the Multi-Application Solar Telescope (MAST) operated by the 
Udaipur Solar Observatory, Physical Research Laboratory, Dept. of Space, Govt. of India. We are thankful to 
the MAST team for providing the data used in this work. We are also thankful for the support being provided 
by Udaipur Solar Observatory/Physical Research Laboratory, Udaipur. S.P.R. acknowledges support from the 
Science and Engineering Research Board (SERB, Govt. of India) grant CRG/2019/003786. Finally, we are thankful to the reviewers for fruitful
comments and suggestions, which improved the presentation of the
results.

\bibliographystyle{cas-model2-names}


\bibliography{hk.bib}



\end{document}